\documentclass[aps,prl,twocolumn,showpacs,superscriptaddress]{revtex4}
\usepackage{graphicx}

\begin{document}

\title {Magnetic Interactions in the Geometrically Frustrated Triangular
Lattice Antiferromagnet $\rm CuFeO_2$}

\author{F.~Ye}
\affiliation{Neutron Scattering Science Division,
Oak Ridge National Laboratory, Oak Ridge, Tennessee 37831-6393 }

\author{J.~A.~Fernandez-Baca}
\affiliation{Neutron Scattering Science Division,
Oak Ridge National Laboratory, Oak Ridge, Tennessee 37831-6393 }
\affiliation{Department of Physics and Astronomy,
The University of Tennessee, Knoxville, Tennessee 37996-1200 }

\author{R.~S.~Fishman}
\affiliation{Materials Science and Technology Division, Oak Ridge
National Laboratory, Oak Ridge, Tennessee 37831-6032 }

\author{Y.~Ren}
\affiliation{Experimental Facilities Division, Argonne National
Laboratory, Argonne, Illinois 60439}

\author{H.~J.~Kang}
\affiliation{NIST Center for Neutron Research, National Institute
of Standards and Technology, Gaithersburg, Maryland 20899-8562 }
\affiliation{Department of Materials Science and Engineering,
University of Maryland, College Park, Maryland 20742 }

\author{Y.~Qiu}
\affiliation{NIST Center for Neutron Research, National Institute
of Standards and Technology, Gaithersburg, Maryland 20899-8562 }
\affiliation{Department of Materials Science and Engineering,
University of Maryland, College Park, Maryland 20742 }

\author{T.~Kimura}
%\affiliation{Los Alamos National Laboratory, Los Alamos, New Mexico
%87545}
\affiliation{Division of Materials Physics, Graduate School of
Engineering Science, Osaka University, Toyonaka, Osaka 560-8531,
Japan
}

\date{\today}

\begin{abstract}
The spin-wave excitations of the geometrically frustrated triangular
lattice antiferromagnet (TLA) $\rm CuFeO_2$ have been measured using
high resolution inelastic neutron scattering.  Antiferromagnetic
interactions up to third nearest neighbors in the $ab$ plane ($J_1$,
$J_2$, $J_3$, with $J_2/J_1 \approx 0.44$ and $J_3/J_1 \approx
0.57$), as well as out-of-plane coupling ($J_z$, with $J_z/J_1
\approx 0.29$) are required to describe the spin-wave dispersion
relations, indicating a three-dimensional character of the magnetic
interactions. Two energy dips in the spin-wave dispersion occur at
the incommensurate wavevectors associated with multiferroic phase,
and can be interpreted as dynamic precursors to the magnetoelectric
behavior in this system.
\end{abstract}

%75.30.Ds spin waves
%75.50.Ee antiferromagnet
%61.12.-q neutron scattering
%\keywords{}
\maketitle

Geometrically frustrated magnetic systems have received considerable
attention in recent years due to the presence of extraordinary
magnetic properties \cite{ramirez99,kimura03,hur04,lottermoser04}.
The delafossite $\rm CuFeO_2$ is of particular interest because of
the discovery of multiferroic phenomena with either the application
of a magnetic field or the substitution of Fe$^{3+}$ with
nonmagnetic Al$^{3+}$ ions
\cite{kimura06,seki06,nakajima07,kanetsuki07}. As a model material
of triangular lattice antiferromagnet (TLA), $\rm CuFeO_2$ forms an
Ising-like 4-sublattice ($\uparrow \uparrow \downarrow \downarrow$)
antiferromagnetic order at low temperature, with spin moment
pointing along the $c$ axis [Figs.~1(b) and 1(c)].  This is contrary to
other TLA where the three spins align at 120$^{\circ}$ from each
other in the basal plane \cite{collins97}.  Despite the Heisenberg
spin of the orbital singlet of Fe$^{3+}$ magnetic ions
($L=0$, $S=5/2$), the metamagnetic transition with magnetic field up
to 60~T \cite{ajiro94} and the successive thermal-induced
magnetic transition \cite{mitsuda98} are well explained with the
two-dimensional (2D) Ising model, where the $\uparrow \uparrow
\downarrow \downarrow$ spin structure is stabilized via the long
range magnetic interactions \cite{takagi95}.  Recent results suggest
that the exchange interaction between adjacent hexagon planes in $\rm
CuFeO_2$ could be equally important as the coupling within the
planes in this seeming 2D system
\cite{petrenko05}. In this Letter, we report high resolution
inelastic neutron scattering (INS) results of the elementary
magnetic excitations in order to clarify the nature of the magnetic
interactions in this TLA. We found that magnetic exchange
interactions up to the third nearest neighbor (NN) have to be
included in order to characterize the dispersion relations in the
basal plane. We also found dispersive excitations perpendicular to the
hexagonal plane indicative of a substantial interlayer coupling. Most
remarkably, we found that the spin-wave dispersion has two dips at
the wavevectors associated with the multiferroic phase. Our
observations reveal that the quasi-Ising like spin order results
from the delicate balance between competing interactions. The static
magnetic order is sensitive to external perturbations (magnetic
field or impurity), and could be transformed into a noncollinear
spin structure that is intimately related to the polar state.

A single crystal $\rm CuFeO_2$ (mass 4.5 g) was grown by the
floating zone technique, and additional small crystals were crushed
to powder for preliminary measurements. The INS measurements were
performed at the Disk Chopper Spectrometer and at the cold neutron
triple-axis spectrometer SPINS at the NIST Center for Neutron
Research. The Disk Chopper Spectrometer measurements on the powder
specimen revealed a magnetic excitation bandwidth between 0.9~meV
and 3.2~meV [Fig.~1(e)]. We therefore constrained our triple-axis
measurements to the energy transfer $E<3.9$~meV.  Although the
crystal undergoes a lattice distortion from hexagonal $R\bar3m$ to a
lower monoclinic $C2/m$ symmetry at low $T$ \cite{ye06,terada06}, we
use the hexagonal notation throughout this paper with lattice
parameter $a = b = 3.03~\rm \AA$ and $c = 17.17~\rm \AA$ for the
sake of simplicity.  The crystal was aligned in the (HHL) scattering
plane. At SPINS vertically focused pyrolytic graphite (PG) crystals
were used as the monochromator, and the final neutron energy was
fixed to $E_f=3.7$~meV.  A BeO filter is placed in the scattered
beam to remove the higher order wavelength contaminations.
Horizontal collimations of open-80'-sample-80'-open were employed.

\begin{figure}[ht!]
\includegraphics[width=2.9in]{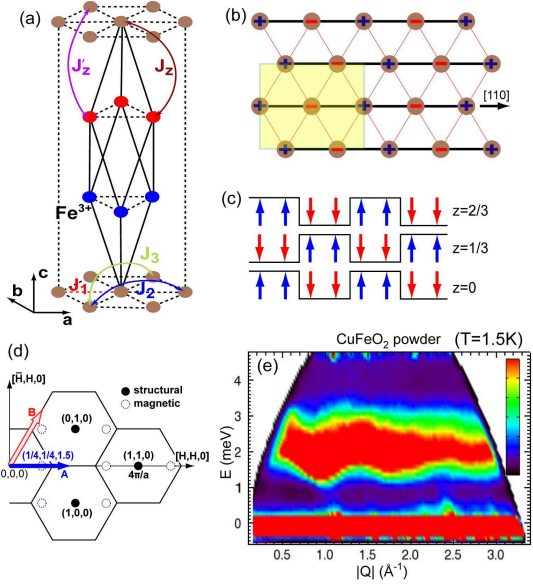}
\caption{\label{fig:structure} (Color online) (a) Schematics of the
crystal structure and the relevant magnetic exchange interactions
between neighboring Fe$^{3+}$ ions. (b) The ground state spin
configuration in the hexagonal plane. The magnetic unit cell is
shown by the (yellow) shaded area. (c) The $\uparrow \uparrow
\downarrow \downarrow$ spin structures are coupled
antiferromagnetically along the $c$ axis. (d) Schematic diagram of
the $(H,K,0)$ scattering plane in reciprocal space.  The structural
Bragg peaks are represented in solid symbols, the magnetic Bragg
peaks displayed in open symbols are projections into that scattering
plane.  (e) Intensity contour map of the inelastic measured using
the time-of-flight Disk Chopper Spectrometer on a powder specimen
(mass $\approx$ 10 g), with an incident neutron wavelength of
3.6~$\rm \AA$.}
\end{figure}

Given the spin structure shown in Figs.~1(b) and 1(c), the magnetic
Bragg peaks appear at $(0.25,0.25,3/2)$ and equivalent positions.
We first studied the magnetic excitations within the hexagonal plane
along the $[H,H,3/2]$ direction [the solid arrow in Fig.~1(d)].
Figure 2(a) displays representative energy scans at constant
wavevector transfer $\vec{Q}$s.  The scattering profiles are highly
structured. In all cases, we identified the lower lying excitation
as the spin-wave branch in the $[H,H,3/2]$ direction, while the
peaks at higher energies originate from unintended scans along the
open arrow shown in Fig.~1(d) due to the crystal hexagonal domains
\cite{twinning}. We notice that the
dispersion does not have a minimum at the magnetic Bragg point
$(0.25,0.25,3/2)$. Instead, two minima appear at symmetric positions at
$\vec{Q}_{m1}=(0.21,0.21,3/2)$ and $\vec{Q}_{m2}=(0.29,0.29,3/2)$,
with a energy gap $\Delta=0.9$~meV, in agreement with the data
reported by Terada {\it et al.}~\cite{terada03}. The
scattering profiles along the $[0,0,L]$ direction, on the other
hand, show much cleaner features due to the absence of twinned
domains. The magnetic excitations along this direction  are clearly
dispersive, ranging in energy from around 1.3~meV at the zone center
to 2.5~meV at the zone boundary. This indicates that the previous
treatment of $\rm CuFeO_2$ as a 2D TLA is an oversimplification.

\begin{figure}[ht!]
\includegraphics[width=3.0in]{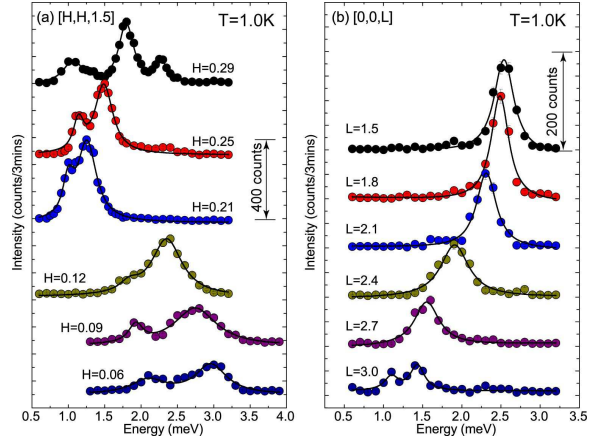}
\caption{\label{fig:qscans} (Color online) Representative spin-wave
excitation scans at $T=1.0$~K. Constant wavevector ($\vec{Q}$) scans
along (a) the $[H,H,3/2]$ direction and (b) the $[0,0,L]$ direction.
Solid lines are Lorentzian fits. Data are shifted for clarity.
}
\end{figure}

In order to obtain an expression of the dispersion relation for $\rm
CuFeO_2$, we start with the generic Heisenberg Hamiltonian:
\begin{equation}
H=-\frac{1}{2}\sum_{i,j} J^{\parallel}_{i,j} \vec{S_i}\cdot\vec{S_{j}} 
-\frac{1}{2}\sum_{i,j} J^{\perp}_{i,j}\vec{S_i}\cdot\vec{S_j} 
- D\sum_{i} S^{2}_{iz},
\end{equation}
where $\sum_{i,j}$ indicates summation over pairs of spins.
$J^{\parallel}_{i,j}$ is the in-plane exchange interactions ($J_1$,
$J_2$, $J_3$ and higher order NN coupling as defined in Fig.~1),
$J^{\perp}_{i,j}$ is the interactions between adjacent planes ($J_z$
and $J_z^{\prime}$) \cite{stacking}. $D$ is the single-ion
anisotropy. Using a Holstein-Primakoff transformation and a $1/S$
expansion, we obtain the analytic form of the spin-wave dispersion
relation: 
\begin{equation}
\varepsilon^2({\bf k}) = A^2-B^2\pm [(C^2-C^{*2})^2 +4|AC-BC^*|^2]^{1/2},
\end{equation}
where
\begin{eqnarray}
& A = &2S[D-J_1+J_2(1-\cos k_y \sqrt{3}a)-J_3(1+\cos 2k_xa)\nonumber \\ 
& & - J_z+2{J_z'} (1-\cos k_zd\cos k_xa)], \nonumber \\
&B = &-2S[(J_1+2J_3 \cos k_y \sqrt{3}a) \cos k_xa + J_z \cos k_zd], \nonumber \\
&C = &-2S\cos (k_y \sqrt{3}a/2) [J_1 e^{ik_xa/2}+J_2 e^{-3i k_x a/2}] \nonumber \\
& & -4S {J_z'}\cos k_zd \cos (k_y \sqrt{3}a/2) e^{-i k_x a/2}. \nonumber
\end{eqnarray}
Here, $k_x$, $k_y$, $k_z$ are the wavevectors along the $[110]$,
$[\bar{1}10]$, and $[001]$ directions respectively, and $d=c/3$ is
the spacing between adjacent hexagonal planes.

\begin{figure}[ht!]
\includegraphics[width=3.2in]{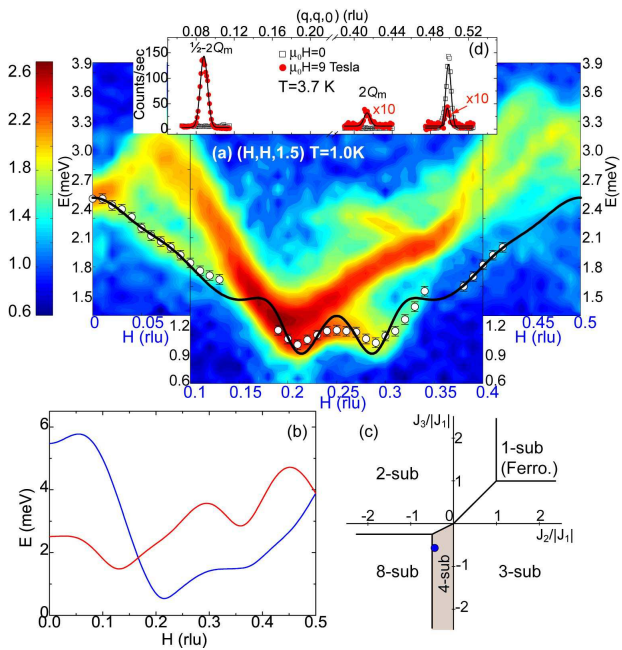}
\caption{\label{fig:HH0scans}
(Color online) (a) Spin-wave dispersion along the $[H,H,3/2]$
direction. Points are experimental data and the solid line denotes a
fit to the model described in the text.  Only the data from the
lowest branch are included in the analysis. The error bars represent
one standard deviation. (b) Predicted dispersion relations along the
open arrow in Fig.~1(d) (contribution from the twinning domains).
(c) Phase diagram from Ref.~[\onlinecite{takagi95}]. The solid
symbol is the value obtained from our measurements. (d) The
structural superlattice peaks appear at $\vec{Q}_L=2\vec{Q}_m$ and
$\vec{Q}_L=1/2-2\vec{Q}_m$ (due to exchange striction) when the
incommensurate spin order occurs at $\mu_0 H=9$~Tesla.
}
\end{figure}

Along the high symmetry $[001]$ direction, the lower branch of the
spin-wave dispersion relation described in Eqn.~(2) reduces to the
much simpler form:
\begin{eqnarray}
& \varepsilon^2(k_z)=  	& 4S^2[(D-G)(D-G+F)+	\nonumber\\
&			& FG\cos{k_zd}-G^2\cos^2{k_zd}],
\end{eqnarray}
where $F=2(J_2-2J_3)$ and $G=J_z-2J_z'$. The bandwidth of
$\varepsilon^2(k_z)$ is mainly controlled by the out-of-plane
interactions $J_z$ and $J_z'$ [{\it i.e.}
$\varepsilon^2(1.5)-\varepsilon^2(3.0)=-8S^2FG$], while the
single-ion anisotropy term $D$ is determined by fitting the
experimental data to the reduced dispersion relation Eq.~(3). We
obtained $DS=0.17$~meV, which is larger than the estimated value of
$DS=0.05$~meV from magnetization measurement \cite{petrenko05}.  The
solid lines in Fig.~3(a) and Fig.~4 represent the global fits to the
observed spin-wave energies along $[H,H,3/2]$, $[0,0,L]$ and
$[q,q,L]$ (with $q=$ 0.21 and 0.25) using the model described above.
Long-range antiferromagnetic interactions (up to $J_3$) are found
($J_1S=-1.14$~meV, $J_2S=-0.50$~meV, and $J_3S=-0.65$~meV) to
satisfactorily describe the in-plane dispersion relation
\cite{distortion}.  The inclusion of higher order interactions
($J_4S$) in our analysis did not improve the fits. Only
one NN exchange interaction ($J_zS=-0.33$~meV) is necessary to
characterize the spin-wave dispersion out of the basal plane, the
addition of a second out-of-plane exchange interaction also yields a
small value ($J_z'S=0.01$~meV) with no significant improvement of
the fit. 

\begin{figure}[ht!]
\includegraphics[width=2.9in]{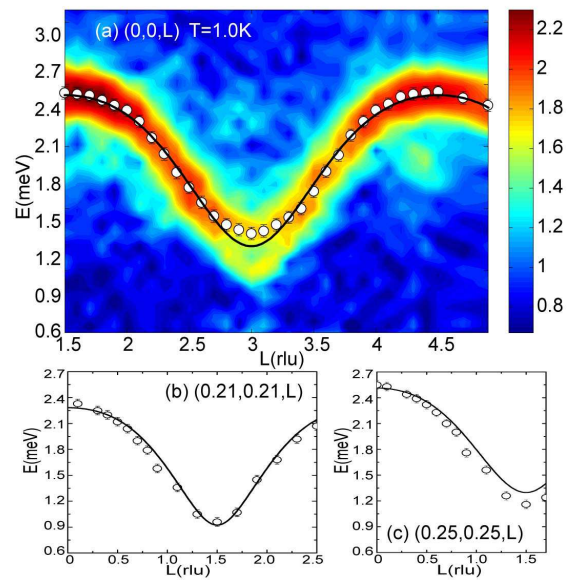}
\caption{\label{fig:scan00L}
(Color online) Spin-wave dispersions along the (a) $[0,0,L]$, (b)
$[0.21,0.21,L]$, and (c) $[0.25,0.25,L]$ direction.  Points are
experimental data and the solid line are fits to the model described
in the text.  Only the data at the lowest branch are presented in
(b) and (c).
}\end{figure}

We now compare our fitting parameters with the earlier theoretical
calculations. The phase diagram of the 2D Ising spin TLA has been
investigated by Takagi and Mekata using Monte Carlo simulations
\cite{takagi95}. They found that the collinear 4-sublattice
structure is only stable with fairly large $J_2$ and $J_3$
contributions. The $\uparrow \uparrow \downarrow \downarrow$ phase
is bound within a narrow range of $-0.5<J_2/|J_1|<0$ and
$J_3/J_2<0.5$ [the shaded area in Fig.~3(c)]. Our result, with
$J_2/J_1=$0.44 and $J_3/J_1$=0.57, is positioned within this shaded
region near the boundary where the 4-sublattice state coexists with
the 8-sublattice state \cite{takagi95}. The phase diagram is not
affected by the inclusion of the single-ion anisotropy $D$, as this
term uniformly shifts the energy at each spin site.  Similarly, this
phase diagram remains unchanged with respect to the NN out-of-plane
exchange coupling $J_z$, since the total energy stays invariant
regardless of the detailed in-plane magnetic structure, providing there
is a negligible $J_z^\prime$.  The agreement between our estimation
of the in-plane exchange constants and those predicted by Takagi and
Mekata implies that the mean field approach can effectively
explain the emergence of the long-range magnetic order. 

Another important result is the observation of two energy dips at
$\vec{Q}_{m1}$ and $\vec{Q}_{m2}$ in symmetric positions around the
commensurate magnetic Bragg point $\tau=(0.25,0.25,3/2)$
characteristic of the 4-sublattice phase. Remarkably these are
precisely the same wavevectors associated with the incommensurate
magnetic order of $\rm CuFeO_2$ under a magnetic field \cite{note},
and are dynamic precursors to the multiferroic phase.  By decreasing
the single-ion anisotropy $D$ without modifying the exchange
constant $J_i$'s, one can essentially close the spin gap $\Delta$,
while the characteristic $\vec{Q}_m$ related to the energy minimum
remain unchanged. In any antiferromagnet, the spin-wave modes are
linearly split in a magnetic field and a critical value of
$H_c(T)=\Delta/g \mu_B$ will destroy the local stability of the
antiferromagnetic state. For $\Delta=0.9$~meV this critical value is 
7.7~Tesla, which is in agreement with the experimental
observation that incommensurate magnetic order appears for $\mu_0
H>7.0$~Tesla \cite{kimura06}. The softening of the magnetic dynamics
achieved by applying a magnetic field reveals a close connection
between the breakdown of the $\uparrow \uparrow \downarrow
\downarrow$ spin structure and the formation of a new type of
magnetic order. The local stability of the antiferromagnetic state
can also be destroyed by chemical substitution. A polar state has
been recently discovered in the doped $\rm CuFe_{1-x}Al_{x}O_2$ in
zero magnetic field possessing an incommensurate spin structure with
the same $\vec{Q}_m$ \cite{nakajima07,kanetsuki07}.  

There remain open questions in this TLA system. For instance, why
are the exchange interactions up to $J_3$ required to stabilize the
collinear spin structure? Although the slight increase in bonding
angle of Fe-O-Fe along the $[110]$ direction [Fig.~1(b)] below the
N\'{e}el temperature  might explain a dominant antiferromagnetic
interaction $J_1$, there is no obvious justification for the
existence of comparable AF interactions between higher order
neighboring spins.  Second, the microscopic origin of the anisotropy
gap $\Delta$ is yet to be understood. One would expect the magnetic
properties should be isotropic since Fe$^{3+}$ is an orbital
singlet. One possible explanation might be related to the lattice
distortion accompanied with the magnetic order which induces an
anisotropy between the Fe-3$d$ and O-2$p$ hybridization as well as
the charge transfer between them \cite{terada07}. Finally, the
excitation data are not completely characterized by the Heisenberg
model described in Eq.~(1).  Figure~3(b) shows the expected
dispersion curves from the twinned magnetic domains. Only some
features agree with the experimental data. Asymmetric dispersion
curves appear with local minima located at $\vec{Q}=(0.22,0.22,3/2)$
and $(0.36,0.36,3/2)$.  However, the magnetic gap and overall energy
bandwidth do not agree with the observations. 

In summary, the spin-wave spectrum of the geometrically frustrated
TLA $\rm CuFeO_2$ has been mapped out using INS. We observed a
considerably dispersive excitation along the $c$ axis, which
reflects the 3D character of the magnetic interactions. We have also
determined the relevant spin Hamiltonian parameters, confirming that
the next NN and the third NN interactions are required to explain
the spin dynamics. Finally, the local minimum of the dispersion
curves reveal the dynamic precursory to the multiferroic phase.

We are grateful to D. I. Khomskii, S. Mitsuda and N. Terada for
useful discussions. ORNL is supported by U.S. DOE under Contract No.
DE-AC05-00OR22725 with UT/Battelle LLC.This work utilized facilities
supported in part by the National Science Foundation under Agreement
No. DMR-0454672.


\begin{references}
\bibitem{twinning} The single crystal has three equally populated
magnetic domains in the basal plane. We simultaneously probed the
excitations from the twinning domains which are $60^\circ$ apart in
the reciprocal space. The stronger scattering intensities at higher
energies are consistent with these contributions.
\bibitem{stacking} We use simplified out-of-plane interactions where
Fe ions couple directly to the ones at the top and the bottom.  This
is slightly different from the actual stacking pattern shown in
Fig.~1(a).
\bibitem{distortion} A known lattice distortion ($\Delta a/a
<0.4\%$, Ref.~[\onlinecite{ye06,terada06}]) may cause a small change
in the exchange constants along one direction that stabilizes the
4-sublattice phase along that direction. It is unlikely, that this
will significantly modify the spin-wave dispersion described in this
paper.
\bibitem{note} Using synchrotron x-ray (Ref.~[\onlinecite{ye06}]) 
and neutron scattering [Fig.~3(d)], we found that an incommensurate
lattice modulation with $\vec{Q}_L=(0.91,0.91,0)$ and
$(0.09,0.09,0)$ appears in the multiferroic phase of $\rm CuFeO_2$.
The relationship between $\vec{Q}_L$ and $\vec{Q}_m$ for the
corresponding incommensurate magnetic structure is given by:
$\vec{Q}_L=(1/2,1/2,0)\pm(2q_m,2q_m,0)$ and
$\vec{Q}_m=(q_m,q_m,3/2)$, where $q_m=0.205$. The two-to-one ratio
of the lattice and magnetic structure is due to the strong
magnetoelastic coupling.
\end{references}
\end{document}